\documentstyle[12pt,epsfig]{article}
                                                                                
\title{                                                                         
{\vspace{-3cm} \normalsize                                                      
\hfill \parbox{30mm}{DESY 96-247}}\\[25mm]                                      
Gluinos on the lattice: quenched calculations    \\[8mm]}
\author{G. Koutsoumbas                             \\
Physics Department, National Technical University, \\
Zografou Campus, 15780 Athens, Greece              \\[4mm]
I. Montvay                                         \\
Deutsches Elektronen-Synchrotron DESY,             \\
Notkestr.\,85, D-22603 Hamburg, Germany}

\date{December, 1996}
                                                                                
\newcommand{\be}{\begin{equation}}                                              
\newcommand{\ee}{\end{equation}}                                                
\newcommand{\half}{\frac{1}{2}}

\begin{document}                                                                
\maketitle                                                                      
                                                                                
\begin{abstract} \normalsize                                                    
 As a preparation for the numerical study of the SU(2) gauge theory
 with gluinos, the spectral properties of the fermion matrix and the
 masses of pseudoscalar and scalar states are investigated in the
 quenched approximation.
 The behaviour of the disconnected fermion diagrams on small lattices
 is also studied. 
\end{abstract}                                                                  
\hspace{1cm}                                                                    
                                                                                
\section{Introduction}                                      \label{sec1}
 The most remarkable developments in non-perturbative Quantum Field
 Theory (QFT) in recent years have been triggered by the exact solution
 of certain supersymmetric gauge theories by Seiberg and Witten
 \cite{SEIWIT}.
 These exact results are exploiting the electric-magnetic duality and
 imply rigorous proofs of confinement and chiral symmetry breaking.
 They open the door for a new non-perturbative procedure: after solving
 the theory in a highly symmetric point of the parameter space one can
 calculate the renormalized effective action by deformations or by
 expansions in symmetry breaking parameters.
 (For examples, see ref.~\cite{ASPY,AGDKM}.)

 The conventional approach of non-perturbative QFT based on lattice
 regularization is in some sense opposite.
 There, in the cut-off theory, the symmetries are broken, with the
 important exception of exact local gauge symmetry, and have to be
 restored in the continuum limit by parameter tuning.
 In spite of the great successes of the Seiberg-Witten approach, it
 would be important to establish a connection to lattice regularization
 by investigating supersymmetric gauge theories on the lattice.
 A possible way to do this was put forward some time ago by Curci and
 Veneziano \cite{CURVEN}.
 (For the case of $N=2$ supersymmetric gauge theories considered in
 \cite{SEIWIT} see also \cite{N=2}.)

 The simplest supersymmetric gauge theory is pure SU(2) Yang-Mills
 theory with massless gluinos.
 In case of massive gluinos the non-vanishing gluino mass softly breaks
 supersymmetry.
 Recently some possible fermion simulation algorithms for theories with
 gluinos have been developed and tested \cite{ALGORI,DONGUA}.
 A numerical simulation with dynamical gluinos is in progress
 \cite{STLOUIS,UNQUENCH}.
 As a first preparatory step, so called ``quenched'' studies are useful,
 where the gauge configurations are prepared in pure gauge theory.

 In the present paper we investigate some interesting features of the
 SU(2) gauge theory with gluinos in the quenched approximation.
 The spectral properties of the fermion operator are studied, which
 are important in the class of {\em local bosonic fermion algorithms} 
 \cite{LUSCHER} we shall apply.
 First information about the interesting region of the fermion hopping
 parameter, where hadronic masses are small in lattice units, can be
 obtained from determining the masses of pseudoscalar and scalar bound
 states.
 In our case these are pion-like, sigma-meson-like or eta-meson-like,
 states which are held together by the colour forces among the gluinos
 in the adjoint representation.
 (In case of SU(2) this means colour triplets.)
 Therefore, we shall call them {\em adjoint-pion} (shortly a-pion), 
 {\em adjoint-sigma} (shortly a-sigma) and {\em adjoint-eta} (shortly
 a-eta), respectively.
 For the a-eta state, the computation of the {\em disconnected fermion
 diagrams} is necessary.

 In the next section we report on our quenched calculations carried out
 at the bare gauge coupling parameter $\beta \equiv 4/g^2 =2.3\;$,
 which roughly corresponds to the lower edge of the scaling region in
 pure SU(2) lattice gauge theory with Wilson action.
                                                                                
\section{Numerical results}                                 \label{sec2}

\subsection{Spectral boundaries of the fermion matrix}
 The {\em fermion matrix} $Q$ relevant for gluinos can be defined as
\be  \label{eq01}
Q_{yv,xu} \equiv Q_{yv,xu}[U] \equiv
\delta_{yx}\delta_{vu} - K \sum_{\mu=1}^4 \left[
\delta_{y,x+\hat{\mu}}(1+\gamma_\mu) V_{vu,x\mu} +
\delta_{y+\hat{\mu},x}(1-\gamma_\mu) V^T_{vu,y\mu} \right] \ .
\ee
 Here, as usual, $x,y,\ldots$ denote lattice sites, $u,v,\ldots$ colour
 triplet indices, $\hat{\mu}$ the unit vector in direction $\mu=1,2,3,4$
 and $K$ the hopping parameter determining the gluino mass.
 The gauge link variable in the adjoint representation $V_{x\mu}$ is
 connected to the usual fundamental link variable $U_{x\mu}$ by the
 relation
\be  \label{eq02}
V_{rs,x\mu} \equiv V_{rs,x\mu}[U] \equiv
2 {\rm Tr}(U_{x\mu}^\dagger T_r U_{x\mu} T_s)
= V_{rs,x\mu}^* =V_{rs,x\mu}^{-1T} \ ,
\ee
 if $T_r$ is a group generator (for SU(2) $T_r=\half\tau_r$ is
 proportional to the Pauli matrix $\tau_r$).
 
\begin{figure}
\begin{center}
\epsfig{file=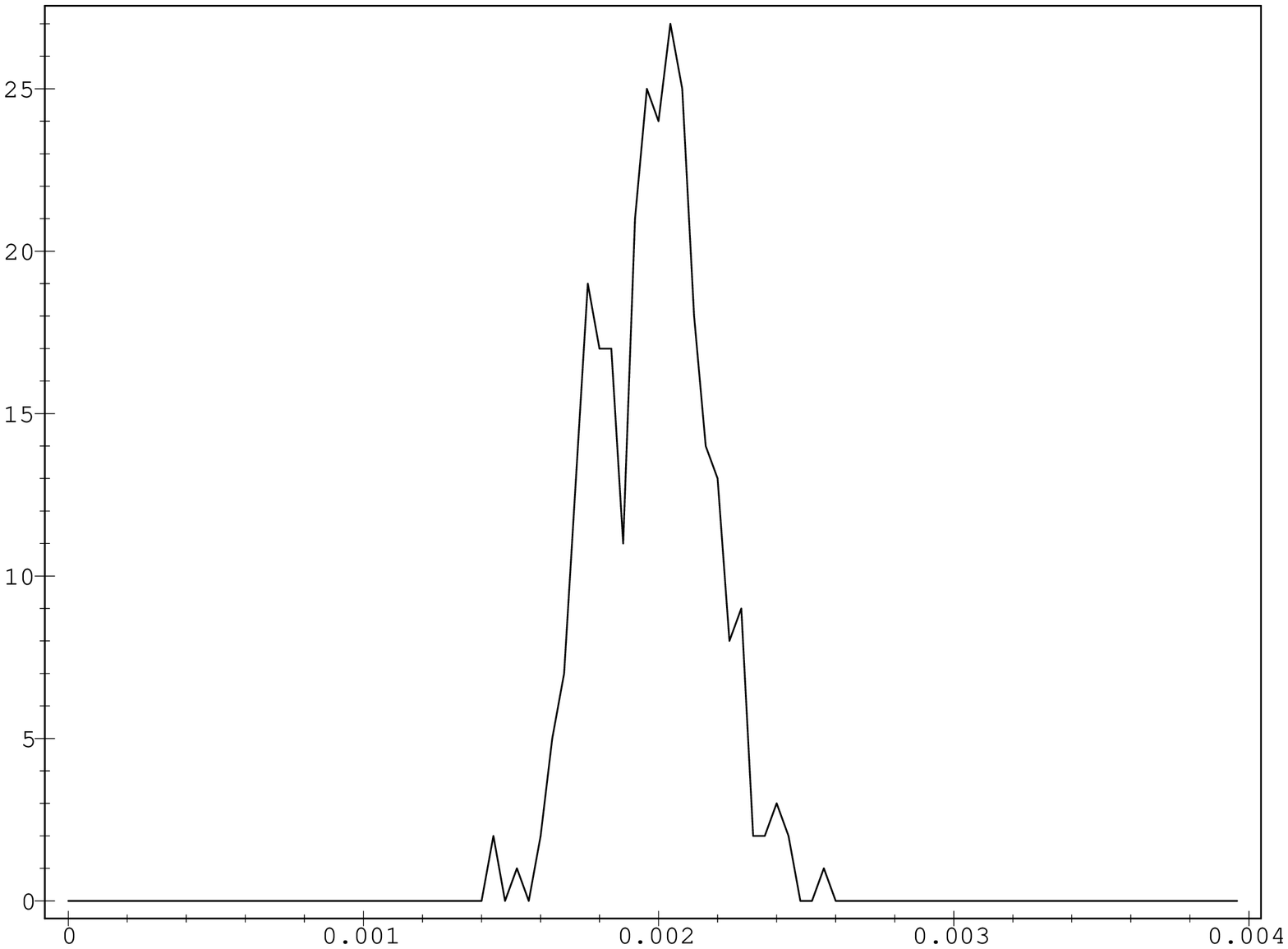,
        width=16.0cm,height=16.0cm,
        bbllx=20pt,bblly=20pt,bburx=590pt,bbury=740pt,
        angle=270}
\parbox{14cm}{\caption{\label{fig01}
 The quenched distribution of the lower spectrum boundary $\rho_{min}$
 of $Q^\dagger Q$ on $8^3 \cdot 16$ lattice at
 $\beta=2.30,\; K=0.20\,$. }}
\end{center}
\end{figure}

\begin{figure}
\begin{center}
\epsfig{file=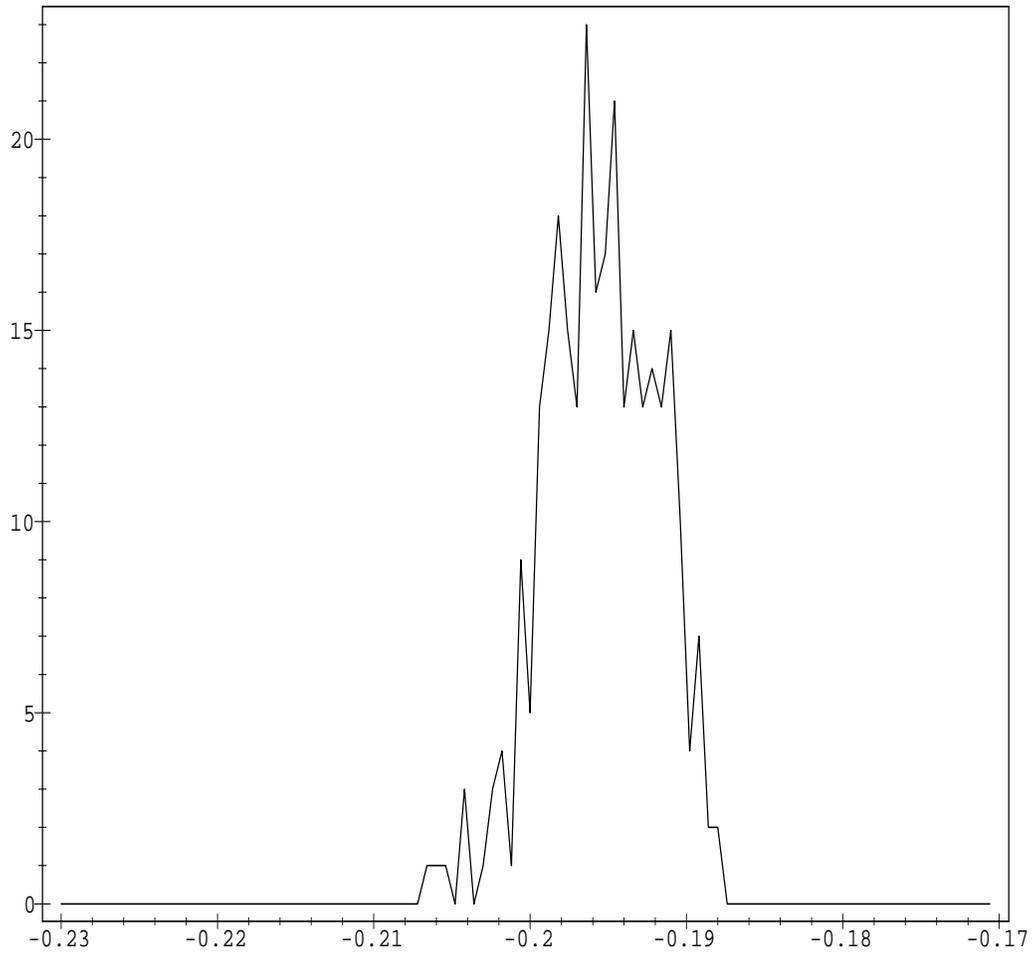,
        width=16.0cm,height=16.0cm,
        bbllx=20pt,bblly=20pt,bburx=590pt,bbury=740pt,
        angle=270}
\parbox{14cm}{\caption{\label{fig02}
 The quenched distribution of the lower spectrum boundary
 $\lambda_{Rmin}$ of $Q_R=(Q+Q^\dagger)/2$ on $8^3 \cdot 16$ lattice at
 $\beta=2.30,\; K=0.20\,$. }}
\end{center}
\end{figure}

 In local bosonic updating algorithms the spectral properties of $Q$
 and $Q^\dagger Q$ are important, because the polynomial approximations
 defining the bosonic action are optimized on the eigenvalues.
 In the two-step local bosonic algorithm \cite{ALGORI} the polynomials
 are optimized in an interval $[\epsilon,\lambda]$ containing the
 eigenvalues $\rho$ of $Q^\dagger Q$.
 For a hermitean matrix as $Q^\dagger Q$ the extremal eigenvalues 
 $\rho_{min} \equiv \min\rho$ and $\rho_{max} \equiv \max\rho$ can
 be obtained by known iterative methods \cite{FADEEV} (for recent work
 on the spectrum of the Wilson-Dirac operator in QCD see \cite{JLSS}).
 In the same way, the extremal eigenvalues of the hermitean matrices
\be  \label{eq03}
Q_R \equiv \half (Q+Q^\dagger) \ , 
\hspace{3em}
Q_I \equiv \frac{1}{2i}(Q-Q^\dagger)
\ee
 can also be determined.
 Since for vectors $v$ normalized to unity we have
\be  \label{eq04}
Q = Q_R +iQ_I \ , 
\hspace{2em}
{\rm Re\,}\langle v |Q| v \rangle = \langle v |Q_R| v \rangle \ ,
\hspace{2em}
{\rm Im\,}\langle v |Q| v \rangle = \langle v |Q_I| v \rangle \ ,
\ee
 the spectral boundaries of $Q$ itself can also be estimated.
 For instance, if $\lambda$ denotes eigenvalues of $Q$ and $\lambda_R$
 eigenvalues of $Q_R$, we have
\be  \label{eq05}
\lambda_{Rmin} \equiv \min \lambda_R = 
\min {\rm Re\,}\langle v |Q| v \rangle \leq \min {\rm Re\,}\lambda \ ,
\ee
 and similarly for the maximum $\lambda_{Rmax}$ of $\lambda_R$ and the
 extrema $\lambda_{Imin,max}$ of $\lambda_I$.
 The numerical simulation results for the quenched averages of these
 extremal eigenvalues at $\beta =2.3$ for $0.180 \leq K \leq 0.205$ on
 $8^3 \cdot 16$ lattices are shown in table \ref{tab01}.
 The distributions of $\rho_{min}$ and $\lambda_{Rmin}$ for $K=0.200$
 are also shown in figures \ref{fig01} and \ref{fig02}, respectively.
 The statistics is based on at least 320 independent gauge
 configurations per hopping parameter value.
 The iteration for the extremal eigenvalues was stopped if the relative
 deviation of the new estimate from the previous one was less than
 $10^{-3}$.
\begin{table}[ht]
\vspace*{-1.0em}
\begin{center}
\parbox{14cm}{\caption{\label{tab01}
 The quenched averages of the extremal eigenvalues defined in the text
 on $8^3 \cdot 16$ lattice at $\beta=2.3\,$.
 The statistical errors are given in parentheses.
}}
\end{center}
\vspace*{-1.0em}
\begin{center}
\begin{tabular}{|c|l|l|l|l|c|}
\hline
  $K$  &  $\rho_{min}$  &  $\rho_{max}$  &
  $\lambda_{Rmin}$  &  $\lambda_{Rmax}$  &
  $\lambda_{Imax}=-\lambda_{Imin}$       \\
\hline\hline
0.180  &  0.00879(2)  &  4.4873(6)  &  -0.0801(1)  &  2.0690(2)  &
0.86357(5)  \\
\hline
0.190  &  0.00434(2)  &  4.764(1)   &  -0.1375(2)  &  2.1287(3)  &
0.91151(9)  \\
\hline
0.200  &  0.00197(1)  &  5.050(1)   &  -0.1958(2)  &  2.1885(3)  &
0.95948(9)   \\
\hline
0.205  &  0.00138(2)  &  5.194(1)   &  -0.2245(2)  &  2.2178(3)  &
0.98373(9)   \\
\hline\hline
\end{tabular}
\end{center}
\end{table}

 As table \ref{tab01} and figure \ref{fig01} show, the lower limit of
 the spectrum of $Q^\dagger Q$ is of the order of $10^{-3}$, and the
 ratio of the largest to smallest eigenvalues is $10^3$-$10^4$.
 This is the range one also encounters in ``unquenched'' numerical
 simulations with dynamical gluinos \cite{UNQUENCH}.
 It is interesting to observe that the lower limit of the spectrum of
 $Q_R$ defined in (\ref{eq03}) is negative in the whole range of the
 hopping parameter considered.
 Although the minimal real part of the eigenvalues of $Q$ might be
 somewhat larger, it is probably also negative.
 This shows that in the relevant hopping parameter range the eigenvalues
 of $Q$ do not remain in the half plane with positive real part, but
 surround zero in the complex plane from all sides.

\subsection{Masses of a-pion and a-sigma}
 The connected contributions to pseudoscalar and scalar bosons are
\be  \label{eq06}
{\rm Tr\,}\{ \gamma_5 Q^{-1}_{yx}\gamma_5 Q^{-1}_{xy} \} \ ,
\hspace{3em}
{\rm Tr\,}\{ Q^{-1}_{yx}Q^{-1}_{xy} \} \ ,
\ee
 respectively.
 These give the correlation functions of the corresponding flavour
 non-singlet mesons in models with at least two (adjoint) fermion
 flavours ($N_a \geq 2$).
 Such states, which we call a-pion and a-sigma, respectively, do not
 exist in the gluino model, which is equivalent to an (adjoint) flavour
 number of $N_a=\half$.
 Nevertheless, the masses of the a-pion and a-sigma are interesting,
 because the hopping parameter region where they become small in
 lattice units is the interesting one, where also other physical colour
 singlet hadrons become light.

 Similarly to the pion in quenched QCD, one expects that the a-pion
 mass-squared vanishes linearly as a function of $1/K$.
 This is the remnant of chiral symmetry for $N_a \geq 2$ in the quenched
 approximation, because the bare quark mass in lattice units is
 $m_q = \half(1/K-1/K_{0})$.
 Here $K_0=K_0(\beta)$ is the hopping parameter where the pion mass
 vanishes.
 Of course, since the gluino corresponds to $N_a=\half$, in the gluino
 model there is no analogous chiral symmetry and the exact value of the
 ``critical'' hopping parameter ($K_{cr}$) where the a-pion has zero
 mass is not directly relevant.
 It is interesting just as an indicator of the {\em scaling region}
 where the continuum limit has to be performed.

\begin{figure}
\begin{center}
\epsfig{file=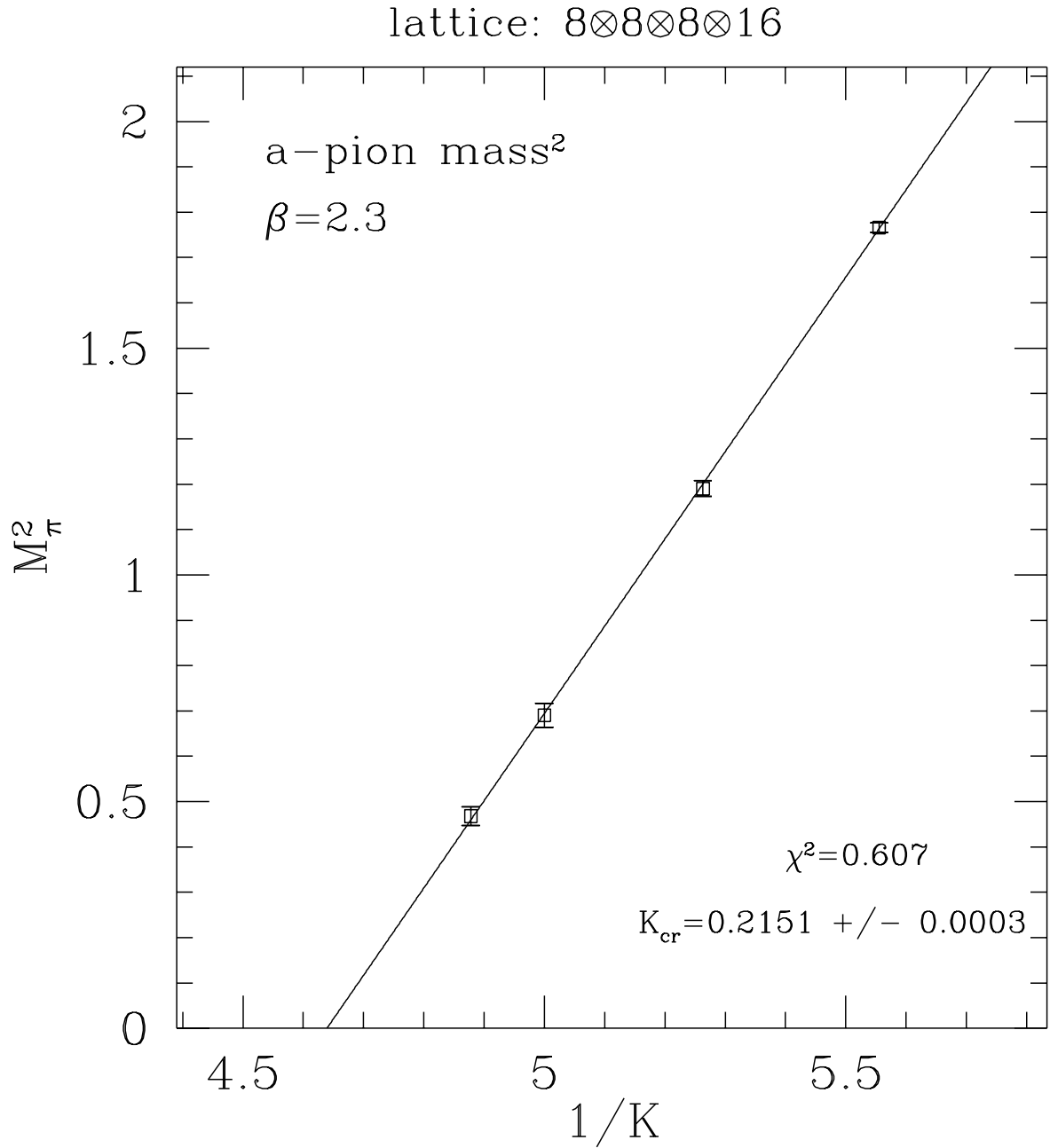,
        width=17.0cm,height=17.0cm,
        bbllx=20pt,bblly=150pt,bburx=600pt,bbury=720pt,
        angle=0}
\parbox{14cm}{\caption{\label{fig03}
 The a-pion mass squared as a function of the inverse hopping
 parameter on $8^3 \cdot 16$ lattice at $\beta=2.3\,$.
 The line is a linear fit used to determine $K_{cr}$.}}
\end{center}
\end{figure}

\begin{figure}
\begin{center}
\epsfig{file=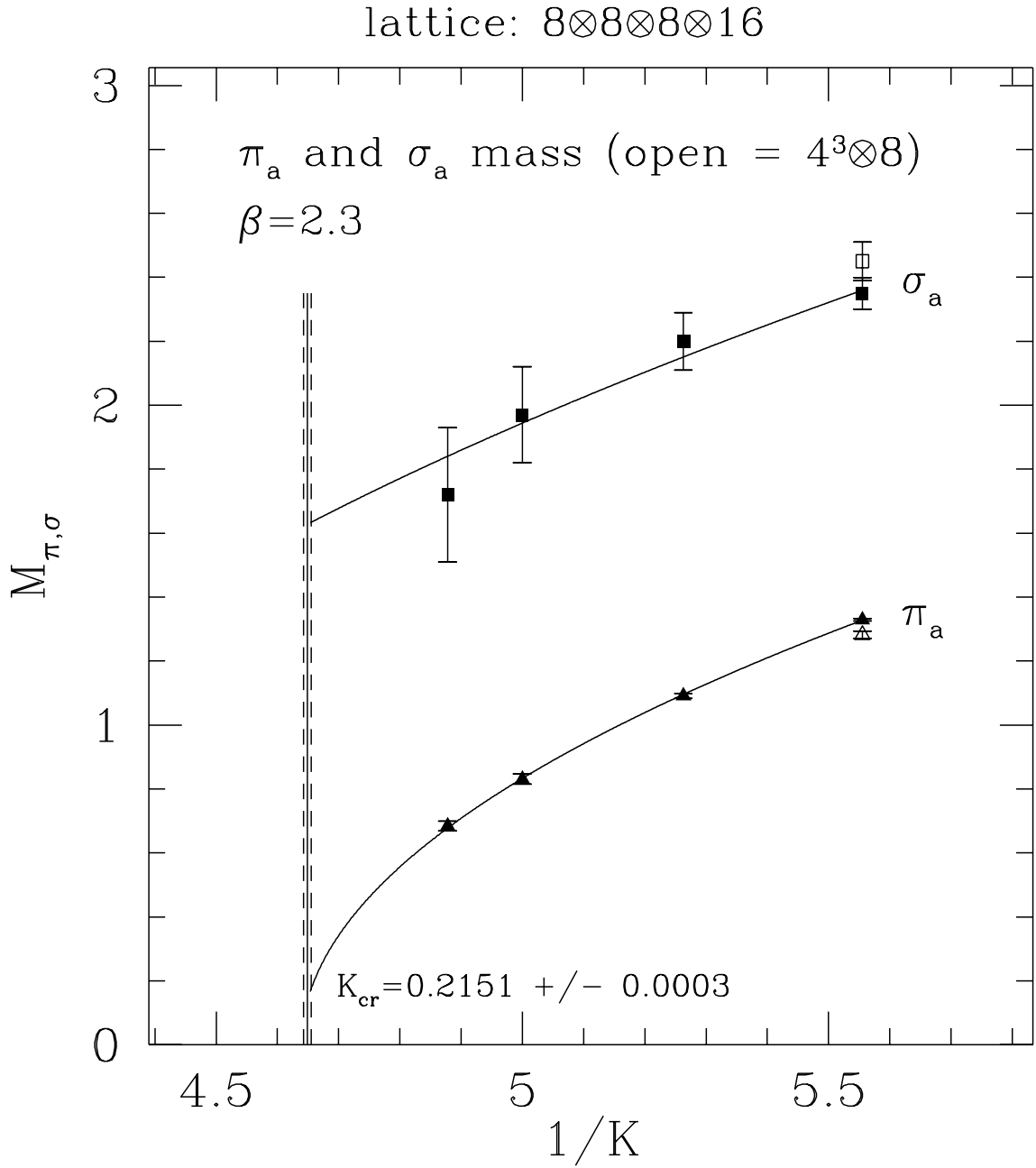,
        width=17.0cm,height=17.0cm,
        bbllx=20pt,bblly=150pt,bburx=600pt,bbury=720pt,
        angle=0}
\parbox{14cm}{\caption{\label{fig04}
 The a-pion and a-sigma mass as a function of the inverse hopping
 parameter on $8^3 \cdot 16$ lattice at $\beta=2.3\,$.
 The open symbols refer to $4^3 \cdot 8$.
 The curves show linear fits for the mass squares.}}
\end{center}
\end{figure}

 The a-pion mass at $\beta=2.3$ as a function of $1/K$ is shown in
 figure \ref{fig03}.
 The linear extrapolation works very well and gives 
 $K_{cr}=0.2151(3)\,$.
 The masses of a-pion and a-sigma are displayed together in figure
 \ref{fig04}.
 As expected, the a-sigma mass is higher and does not seem to vanish
 at $K_{cr}$.
 The results in figures \ref{fig03} and \ref{fig04} were obtained on
 an $8^3 \cdot 16$ lattice.
 The masses were always extracted from timeslice correlations by a fit
\be  \label{eq07}
c_0 + c_1 \left\{ e^{-mt}+e^{-m(T-t)} \right\} \ ,
\ee
 with $T$ being the lattice time extension.
 For comparison, in figure \ref{fig04} at $K=0.180$ also the masses on
 $4^3 \cdot 8$ are included, which do not deviate too much from the 
 corresponding masses on $8^3 \cdot 16$.
 This shows that, not too close to $K_{cr}$, even a $4^3 \cdot 8$
 lattice can give first indications on the mass spectrum. 

\subsection{Disconnected contributions: the mass of a-eta}
 The correlation function of the pseudoscalar meson made out of
 Majorana gluinos can be obtained from the general formulas given
 in ref.~\cite{ALGORI}:
\be  \label{eq08}
\left\langle (\overline{\Psi}_y \gamma_5 \Psi_y)
(\overline{\Psi}_x \gamma_5 \Psi_x) \right\rangle =
\left\langle {\rm Tr\,} \{ \gamma_5 Q^{-1}_{yy} \}
{\rm Tr\,} \{ \gamma_5 Q^{-1}_{xx} \} -
2\,{\rm Tr\,}\{ \gamma_5 Q^{-1}_{yx}\gamma_5 Q^{-1}_{xy} \}
\right\rangle_U \ .
\ee
 Here the index $U$ denotes an expectation value in the path integral
 over gauge variables.
 A similar formula is also valid for scalar mesons, where $\gamma_5$ is
 replaced by the unit Dirac matrix.
 Since in QCD the correlation analogous to (\ref{eq08}), apart from the
 factor 2 in front of the connected contribution, belongs to $\eta$ and
 $\eta^\prime$ mesons, one can call the corresponding boson a-eta. 

 As discussed in ref.~\cite{ALGORI}, the expectation value in
 eq.~(\ref{eq08}) can also be expressed by a ``noisy estimator'' in 
 terms of the $n$ bosonic pseudofermion fields ($\phi$).
 In the present case we have
$$
\left\langle (\overline{\Psi}_y \gamma_5 \Psi_y)
(\overline{\Psi}_x \gamma_5 \Psi_x) \right\rangle =
$$
\be  \label{eq09}
= 4\sum_{j,k=1}^n \left\langle
\left[ (\overline{\phi}^*_{ky}\phi_{ky}) +
(\phi^*_{ky}\overline{\phi}_{ky}) \right]
\left[ (\overline{\phi}^*_{jx}\phi_{jx}) +
(\phi^*_{jx}\overline{\phi}_{jx}) \right] - 2\delta_{jk}\delta_{yx}
(\phi^*_{jx}\phi_{jx}) \right\rangle_{\phi U} \ ,
\ee
 with the notation
\be \label{eq10}
\overline{\phi}_{jx} \equiv \sum_y  (\gamma_5 Q+\mu_j)_{xy} 
\phi_{jy} \ .
\ee
 In the last formula $\mu_j\; (j=1,\ldots,n)$ denote the real parts of
 the roots of the polynomial used in the local bosonic updating
 algorithm.

 The advantage of the noisy estimator in eq.~(\ref{eq09}) is that in
 principle it can be obtained with very little work from the
 pseudoscalar fields during unquenched updating.
 Unfortunately, this does not work in practice: the result is too noisy
 and the signal cannot be obtained in this way \cite{UNQUENCH}.
 What remains is the evaluation of the fermion propagators appearing
 in eq.~(\ref{eq08}).

\begin{figure}
\begin{center}
\epsfig{file=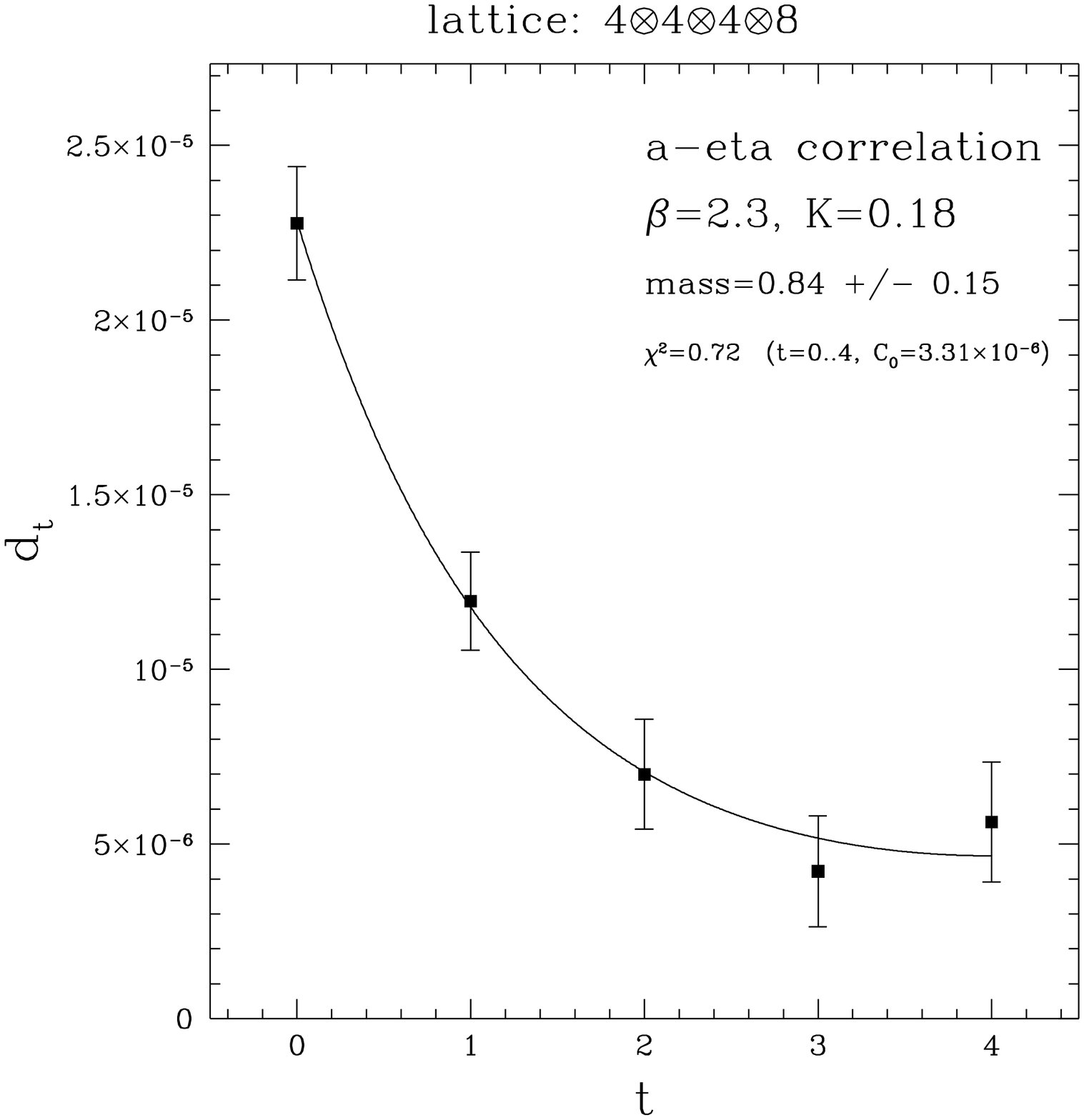,
        width=17.0cm,height=17.0cm,
        bbllx=20pt,bblly=150pt,bburx=600pt,bbury=720pt,
        angle=0}
\parbox{14cm}{\caption{\label{fig05}
 The disconnected a-eta correlation as a function of the timeslice
 distance on $4^3 \cdot 8$ lattice at $\beta=2.3,\; K=0.180\,$.
 The curve is a fit according to (\protect\ref{eq07}) for
 $0 \leq t \leq 4$, giving a mass $m_{disc} = 0.84(15)$.}}
\end{center}
\end{figure}

 The ``connected'' contribution in the second term has been discussed
 in the previous subsection.
 The behaviour of the first ``disconnected'' contribution on 
 $4^3 \cdot 8$ lattice at $\beta=2.3,\;K=0.180$ is shown in figure
 \ref{fig05}.
 Compared to the connected term, the disconnected contribution is two
 to three orders of magnitude smaller.
 On the other hand, the disconnected contribution decreases considerably
 slower as a function of the distance.
 It shows an effective mass of the order of $m_{disc} \simeq 0.8$,
 whereas the effective mass in the connected contribution (the a-pion
 mass) is about $m_{conn} \simeq 1.3\,$.
 This behaviour is qualitatively similar to the one observed in QCD
 \cite{KFMOU}.

 As argued in ref.~\cite{KFMOU}, in the quenched approximation one can
 also extract a ``quenched'' a-eta mass from the ratio of the
 disconnected to connected contributions.
 Denoting the ratio of the two terms at timeslice distance $t$ by
 $R(t)$, one has
\be  \label{eq11}
R(t) = {\rm const.} + t \frac{m_\eta^2}{2m_\pi} \ . 
\ee
 Here $m_\eta$ and $m_\pi$ stand generically for the masses of the
 flavour singlet and flavour non-singlet pseudoscalar mesons,
 respectively.
 In our case we get from (\ref{eq11}) a quenched a-eta mass of about
 $m_{a-eta} \simeq 0.1$, although the linear fit on our small
 $4^3 \cdot 8$ lattice is not very good.
 Of course, we are finally interested in numerical results with
 dynamical gluinos, where the linear behaviour in (\ref{eq11}) is
 irrelevant, and the a-eta mass can be obtained from a usual kind
 of fit of the combination in eq.~(\ref{eq08}) according to
 eq.~(\ref{eq07}).
 The results shown in figure \ref{fig05} were obtained by direct
 evaluation of the necessary inverses on 100 independent
 configurations.
 In fact, for a more precise calculation of the disconnected
 contributions, especially on larger lattices, there are better
 methods known (see ref.~\cite{SESAM}).

 In summary: the quenched calculations give useful information on the
 spectral properties of fermion matrices and the range of hopping
 parameters relevant for the continuum limit, which help in setting up
 numerical simulations with dynamical gluinos.

\vspace{2em}
{\large\bf Acknowledgements}
\newline\vspace{-0.5em}

\noindent  We thank Gernot M\"unster and Dirk Talkenberger for helpful
 discussions.


\newpage
\newpage
\newpage
\newpage
\newpage


\newpage
\begin{thebibliography}{99}
%
\bibitem{SEIWIT}
N. Seiberg, E. Witten,
Nucl. Phys. \underline{B426} (1994) 19;      
ERRATUM ibid. \underline{B430} (1994) 485; \\
Nucl. Phys. \underline{B431} (1994) 484.
%
\bibitem{ASPY}
O. Aharony, J. Sonnenschein, M.E. Peskin, S. Yankielowicz,
Phys. Rev. \underline{D52} (1995) 6157.
%
\bibitem{AGDKM}
L. \'Alvarez-Gaum\'e, J. Distler, C. Kounnas, M. Mari\~no, 
Int. Journ. Mod. Phys. \underline{A11} (1996) 4745. 
%
\bibitem{CURVEN}
G. Curci, G. Veneziano,
Nucl. Phys. \underline{B292} (1987) 555.
%
\bibitem{N=2}
I. Montvay,
Phys. Lett. \underline{B344} (1995) 176;
Nucl. Phys. \underline{B445} (1995) 399.
%
\bibitem{ALGORI}
I. Montvay,
Nucl. Phys. \underline{B466} (1996) 259.
%
\bibitem{DONGUA}
A. Donini, M. Guagnelli,
Phys. Lett. \underline{B383} (1996) 301.
%
\bibitem{STLOUIS}
I. Montvay,
hep-lat/9607035, to appear in 
{\em Proceedings of the Lattice 96 Conference in St. Louis}.
%
\bibitem{UNQUENCH}
G. Koutsoumbas, I. Montvay, A. Pap, K. Spanderen, D. Talkenberger, 
in preparation.
%
\bibitem{LUSCHER}
M. L\"uscher,
Nucl. Phys. \underline{B418} (1994) 637.
%
\bibitem{FADEEV}
D.K. Fadeev, V.N. Fadeeva,
{\em Numerical Methods of Linear Algebra}, (in German),
M\"unchen, 1964.
%
\bibitem{JLSS}
K. Jansen, C. Liu, H. Simma, D. Smith,
hep-lat/9608048, to appear in 
{\em Proceedings of the Lattice 96 Conference in St. Louis}.
%
\bibitem{KFMOU}
Y. Kuramashi, M. Fukugita, H. Mino, M. Okawa, A. Ukawa,
Phys. Rev. Lett. \underline{72} (1994) 3448.
%
\bibitem{SESAM}
SESAM Collaboration (N. Eicker et al.),
hep-lat/9608040.
%
\end{thebibliography}
\end{document}